\documentclass{aa}  
\usepackage{graphicx}
\usepackage{txfonts}
\usepackage{natbib}
\begin{document}
   \title{The very nearby M/T dwarf binary SCR 1845-6357
   \thanks{Based on observations collected at the European Southern Observatory, Chile, through proposal 276.C-5038(A).}}


   \author{M. Kasper\inst{1}
          \and B.A. Biller\inst{2}
          \and A. Burrows\inst{2}
          \and W. Brandner\inst{3}
          \and J. Budaj\inst{2,4}
          \and L.M. Close\inst{2}
          }

   \offprints{M. Kasper, \email{mkasper@eso.org}}

   \institute{European Southern Observatory (ESO), Karl-Schwarzschild-Str. 2, D-85748 Garching
         \and Steward Observatory, University of Arizona, 933 North Cherry Avenue, Tucson, AZ 85721
         \and Max-Planck-Institut f\"ur Astronomie, K\"onigstuhl 17, D-69117 Heidelberg
         \and Astronomical Institute, Tatranska Lomnica, 05960, Slovak Republic
             }

   \date{Received xx, 2007; accepted xx, xx}

 
  \abstract
   {The recently discovered star SCR 1845-6357 (hereafter SCR 1845) is the first late M/T dwarf binary discovered. SCR 1845 is a particular object due to its tight orbit (currently around 4 AU) and its proximity to the Sun (3.85\,pc).}
   {We present spatially resolved VLT/NACO images and low resolution spectra of SCR 1845 in the J, H and K near-infrared bands. Since the T dwarf companion, SCR 1845B, is so close to the primary SCR 1845A, orbital motion is evident even within a year. Following the orbital motion, the binary's mass can be measured accurately within a decade, making SCR 1845B a key T-dwarf mass-luminosity calibrator. The NIR spectra allow for accurate determination of spectral type and also for rough estimates of the object's physical parameters.}
   {The spectral type of SCR 1845B is determined by direct comparison of the flux calibrated JHK spectra with T dwarf standard template spectra and also by NIR spectral indices obtained from synthetic photometry. Constrained values for surface gravity, effective temperature and metallicity are derived by comparison with model spectra.}
   {Our data prove that SCR 1845B is a brown dwarf of spectral type T6 that is co-moving with and therefore gravitationally bound to the M8.5 primary. Fitting the NIR spectrum of SCR 1845B to model spectra yields an effective temperature of about 950\,K and a surface gravity $\log{g}=5.1$ (cgs) assuming solar metallicity. Mass and age of SCR 1845B are in the range 40 to 50 Jupiter masses and 1.8 to 3.1 Gyr.}

   \keywords{astrometry -- stars: late-type -- stars: low mass, brown dwarfs -- stars: binaries: general}

   \maketitle
%

\section{Introduction}

After decades of little change in the number of known stellar systems in the solar neighborhood, large plate digitization efforts such as the Digitized Sky Survey and SuperCOSMOS \citep{hambly01} as well as surveys like the Two Micron All Sky Survey \citep[2MASS,][]{skrutskie06} have led to the discovery and first order characterisation of numerous previously unknown nearby low mass stars \citep[e.g.][]{hambly04}. Many of these stars are very nearby and of low intrinsic luminosity, so they are ideal targets to search for low mass companions, since even a close companion would appear with a reasonably wide separation on the sky and reasonably bright when compared to the primary. 

Especially interesting are multiple systems with one component being a main sequence star and at least one other component having a mass below the hydrogen-burning limit, and hence residing in the domain of brown dwarfs. 
Since all components are expected to be coeval and of the same chemical composition, age and metallicity of the main sequence primary determined by well established methods also applies to the brown dwarf companion. Furthermore, sufficiently tight binary systems would allow one to directly measure the system mass by determining the orbit and even to measure the individual components' masses via monitoring of radial velocities. Hence, the measured luminosity of such a companion can be used to calibrate effective temperature and surface gravity figures of evolutionary models and provide the required benchmarks.

One such precious benchmark object, Eps~Indi~B, was resolved into a binary T dwarf \citep{mccaughrean04} orbiting the K4.5V primary Eps~Indi~A during the commissioning of the Simultaneous Differential Imager \citep[SDI,][]{lenzen04,close05} at the VLT. Another T dwarf companion, SCR~1845B, was recently found by \citet{biller06} around the M8.5 primary SCR~1845A using the same instrument. SCR 1845 was first reported in \citet{hambly04} to be a very red, nearby star at a distance based on plate and 2MASS magnitudes of $3.5\pm0.7$\,pc. A little later, \citet{deacon05} measured a trigonometric parallax, $282\pm23$\,mas, based on SuperCOSMOS photographic plates. This estimate was refined by \citet{henry06} who measure a parallax of $259.45\pm1.11$\,mas ($3.85\pm0.02$\,pc), consistent with all previous estimates. Hence, at its distance of 3.85 pc, SCR 1845 is only slightly further away than Eps Indi and the 24th closest star system to Earth known today. This proximity makes SCR~1845B readily accessible to observations despite its small projected separation of only about 4~AU to the primary and its intrinsic faintness.

It is now widely accepted that, besides $T_\mathrm{eff}$ (a function of bolometric luminosity and radius which are both a function of age), surface gravity (a function of mass and radius) and metallicity have a paramount impact on the spectral energy distribution of late T dwarfs \citep{gorlova03,liu07}. Binary systems that allow us to directly measure as many of these parameters as possible are therefore the key for benchmarking the models and understanding brown dwarf evolution. The mass of SCR~1845 can readily be obtained in the longer term following the orbit which has a period of the order of 20 years \citep{henry06}. The individual components' masses can further be determined from additional relative radial velocity measurements. The determination of metallicity and age of late M dwarfs is still rather difficult but a field of active research. Recent methods to derive i) the metallicity of early to mid M dwarfs via atomic \citep{woolf05} and molecular \citep{woolf06} indices as well as via absolute magnitudes \citep{bonfils05}, and ii) the age of M dwarfs from their chromospheric activity \citep{silvestri05} offer interesting perspectives in this respect. 

While \citet{biller06} were able to constrain the spectral type of SCR~1845B to T5.5$\pm$1, the SDI imaging data did not allow them to deduce a more accurate spectral type and physical properties such as mass, age, metallicity, $T_\mathrm{eff}$ and $L_\mathrm{bol}$. This paper presents follow-up NIR photometry and spectroscopy that is used to accurately determine spectral type by direct comparison to the spectra of T dwarf standards and through spectral indices \citep{burgasser06a}, and to further constrain physical properties by direct comparison with modeled brown dwarf spectra \citep{burrows97, burrows06}. In addition, we also applied a novel method \citep{burgasser06b} for estimating $T_\mathrm{eff}$ and surface gravity of T dwarfs in the absence of higher resolution spectra.
 

\section{Observations and Data Reduction}
\begin{figure*}
   \centering
   \includegraphics[width=15cm]{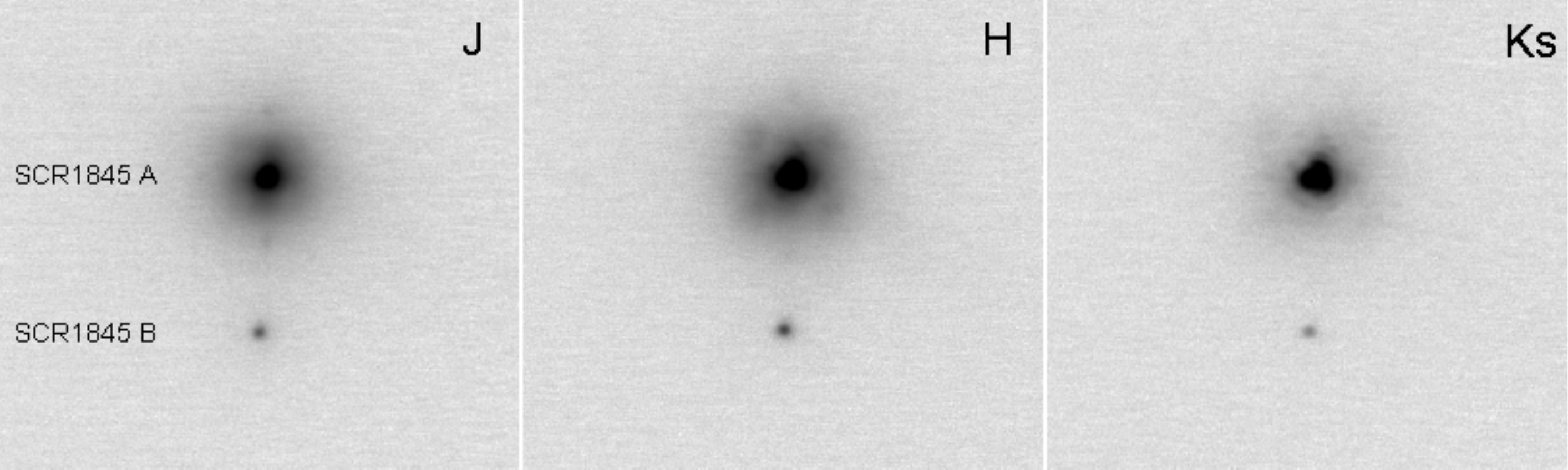}
   \caption{NACO broad band NIR images of the 1\farcs064 binary SCR 1845 with angular resolution (FWHM) of about 67 mas in J-, 64 mas in H-, and 71 mas in K-band. North is up, East is left. Intensities are displayed in a logarithmic scale using identical cut levels in the different bands.}
   \label{fig:jhkima}
\end{figure*}

SCR\,1845 was observed with NACO mounted on UT4 of the VLT on 2 May 2006. The Adaptive Optics used 90\% of the light from the M8.5 primary as a guide star for its infrared wavefront sensor while 10\% of the light were directed towards the scientific camera. In this way, we obtained well corrected images that easily resolve the 1\arcsec{} binary (see Figure\,\ref{fig:jhkima}).

\subsection{Imaging}
For the imaging observations, the S13 camera of NACO was used in the J, H- and Ks near-infrared filters providing a pixel scale of 13.27\,mas per pixel. The target was observed with a standard dithering procedure integrating for one minute on source in each of the bands.

Image data reduction followed standard procedures including sky removal, bad pixel cleaning and flat fielding. Differential photometry and astrometry were done on the reduced images with the PSF fitting DAOPHOT package. The primary was used as the PSF for the photometry.  We used an iterative procedure in order to attain an accurate PSF. With the original binary image as an initial PSF estimate, the secondary was subtracted out of the image in order to produce a second PSF.  This second PSF was then used to attain accurate relative magnitudes for both components of the system differing by less than 3\% from initial estimates. The individual magnitudes for SCR~1845A and B were finally calculated from the relative magnitudes and the integrated 2MASS magnitudes of the unresolved system \citep[J=9.54, H=8.97 and Ks=8.51,][]{henry04}. The error introduced by the difference between NACO and 2MASS instrument response has been calculated from the spectra as described in the next section and has been taken into account. Errors of the astrometric positions were estimated from the standard deviation in the measured separation between the components.

Table\,\ref{tab:photastrom} displays 2MASS photometry and astrometry of SCR 1845A and B. Given the large proper motion of SCR~1845 of 2\farcs444 RA and 0\farcs696 Dec. per year, the relatively small change in relative position between the two components over the one year between the two measurements confirms the result of \citet{montagnier06} and unambiguously proves that SCR 1845 is a bound binary.

\begin{table*}
\begin{minipage}{\linewidth}
\caption{Photometry in the 2MASS system and astrometry of SCR 1845A and B.}
\label{tab:photastrom}
\centering
\renewcommand{\footnoterule}{}  
\begin{tabular}{llllllll}
\hline \hline
Object & Date & JD & J & H & Ks & Separation & PA  \\
\hline
   SCR 1845A & 2 May 2006 & 2453858 & 9.58$\pm$0.024 & 8.99$\pm$0.028 & 8.52$\pm$0.022 & \\
   SCR 1845B & 2 May 2006 & 2453858 & 13.26$\pm$0.024 & 13.19$\pm$0.028 & 13.69$\pm$0.022 & 1\farcs064$\pm$0\farcs0038 &  177$\fdg$2$\pm$0$\fdg$06 \\
    & 28 May 2005\footnote{\citet{biller06}} & 2453519 & & 13.16$\pm$0.28 & & 1\farcs170$\pm$0\farcs003 & 170$\fdg$2$\pm$0$\fdg$13 \\
\hline
\end{tabular}
\end{minipage}
\end{table*}

\subsection{Spectroscopy}
After direct imaging on 2 May 2006, NACO was used in long slit spectroscopic mode to obtain $R\approx400$ spectra in the J and HK spectroscopic filters (modes S54-4-SJ and S54-4-SHK, 2nm/px pixel dispersion, S54 camera with 54.3 mas/pixel). By turning the Nasmyth adaptor rotator, the slit was aligned with the binary axis. Dithering the binary at two positions along the slit, exposures of 15 minutes total in each of the modes were obtained. A telluric calibrator (HIP 95567, spectral type A0V) has been observed in a similar manner just after the object at a similar airmass. A spectrum of the NACO internal Tungsten lamp was recorded for wavelength calibration.

Spectroscopic data reduction followed standard procedures. After sky removal, bad pixel cleaning and flat fielding, the spectra from SCR 1845A and B were extracted and wavelength calibrated. The halo of SCR 1845A at the position of B was estimated by interpolation. The spectra were divided by the telluric calibrator and multiplied by a Kurucz template spectrum of Vega (http://kurucz.harvard.edu/stars/VEGA/) smoothed to the actual resolution of the data in order to remove the hydrogen absorption features of the A0V telluric calibrator. 

Synthetic magnitudes were calculated from the non-flux calibrated spectra by integrating them over the instrument response curve. The ratio between the synthetic magnitudes of SCR 1845A and B was calculated by integrating the spectra over i) the spectral response of the 2MASS system \citep{cohen03}, and ii) the NACO filter response. A comparison between the flux ratios calculated from the same spectra in both photometric systems yields multiplication factors of 1.022 in J, 1.177 in H, and 0.964 in Ks to convert from NACO to 2MASS flux. The rather large correction in H-band originates from the wide NACO filter and the peaked flux distribution of the T dwarf companion in this band, while the M8.5 primary magnitude does not vary much with the assumed photometric system.

\begin{figure}
\centering
\includegraphics[width=\columnwidth]{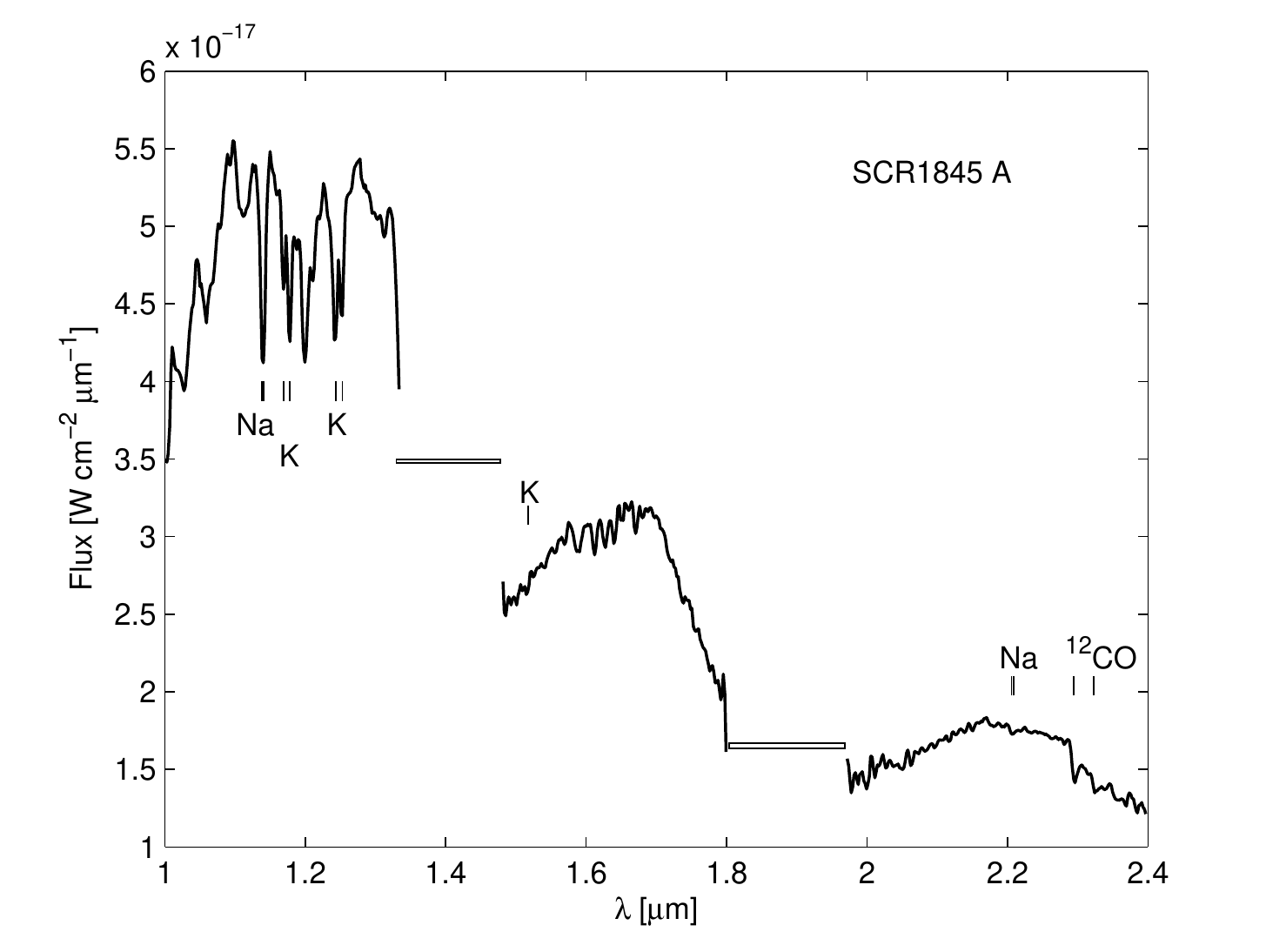}
\caption{NIR spectrum of SCR 1845A.}
\label{fig:jhkspectraA}
\end{figure}

\begin{figure}
\centering
\includegraphics[width=\columnwidth]{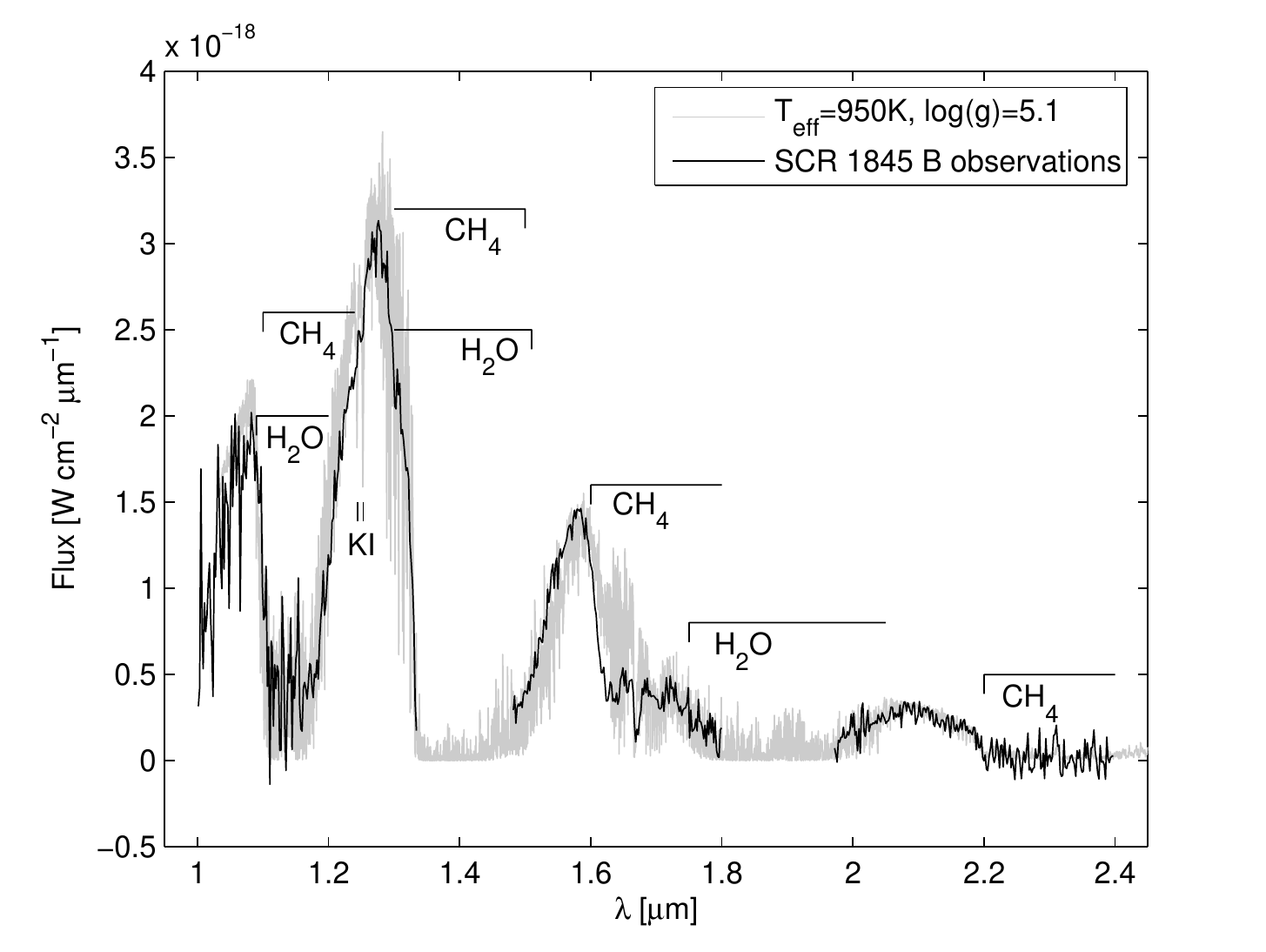}
\caption{NIR spectrum of SCR 1845B. The underlying gray curve shows the best model fit obtained for $T_\mathrm{eff}=950$\,K, $\log{g}=5.1$ (cgs) and solar metallicity.}
\label{fig:jhkspectraB}
\end{figure}

Flux calibration of the spectra was finally established using the calculated 2MASS magnitudes of both components (see table\,\ref{tab:photastrom}) and the 2MASS mag0 flux \citep{cohen03}. Figures\,\ref{fig:jhkspectraA} and \ref{fig:jhkspectraB} show the final NIR spectra of SCR 1845A and B. The figures also indicate the main absorption lines and bands as well as regions of strong telluric absorption.

\section{Results}

\subsection{Spectral Types}
\begin{figure}
\centering
\includegraphics[width=\columnwidth]{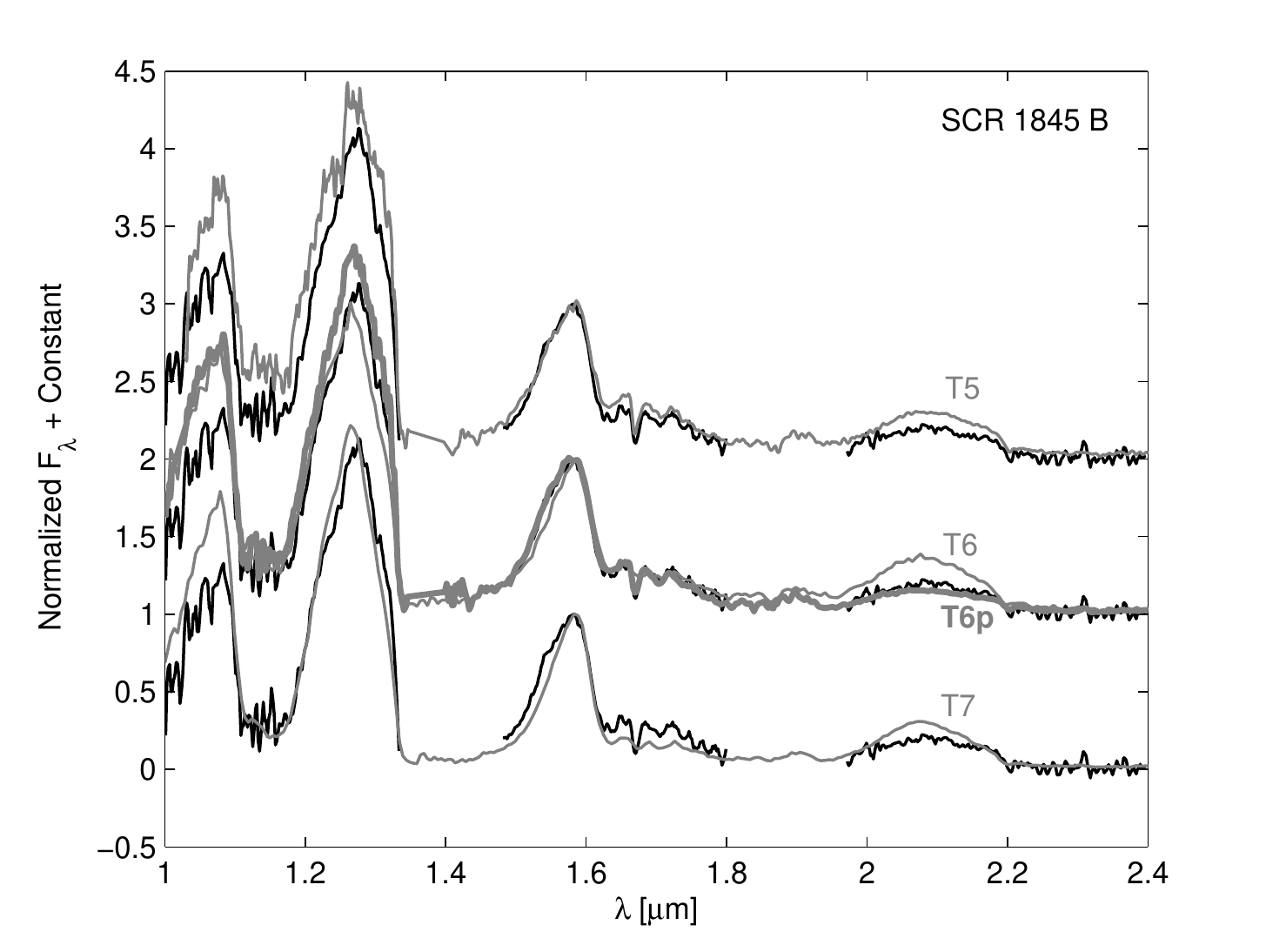}
\caption{Spectrum of SCR 1845B (black line) plotted on top of T5-7 dwarf spectral standards (normalized to the H band).}
\label{fig:SpT}
\end{figure}

The most direct way to determine the spectral type of a star is direct comparison to spectral standards. Figure\,\ref{fig:SpT} plots the NIR spectrum of SCR~1845B (black line) on top of T5-7 dwarf spectral standards \citep{burgasser06a}. The T6 standard SDSS~1624+0029 provides a good fit in the J- and H-bands while the peculiar T6p dwarf 2MASS~0937+2931 better fits the observed K-band flux and leads to a reasonable overall fit. 2MASS~0937+2931 is considered the prototype example of a high surface gravity T dwarf showing the effects of photospheric pressure.

%

In case a complete NIR spectrum does not exist, the spectral type of SCR~1845B can also be determined from spectral indices. Table\,\ref{tab:indices} shows the indices calculated from the spectra over the wavelength ranges for numerator and denominator defined by \citet{burgasser06a}. Three (H$_2$O-J, H$_2$O-H, CH$_4$-H) out of the four classification indices are consistent with spectral type T6. Only the CH$_4$-K index indicates an even later spectral type, but the numerator range of this index (2.215-2.255 $\mu$m) is in a low SNR part of the spectrum and probably not very reliable.

\begin{table}
\caption{Spectral indices of SCR 1845B.}             
\label{tab:indices}      
\centering                          
\begin{tabular}{ccccccc}        
\hline\hline                 
H$_2$O-J & H$_2$O-H & H$_2$O-K & CH$_4$-H & CH$_4$-K & K/J & K/H\\    
\hline                        
   0.155 & 0.297 & 0.461 & 0.286 & 0.097 & 0.101 & 0.221\\      
\hline                                   
\end{tabular}
\end{table}

The NIR spectrum of SCR 1845A agrees with the spectral type M8.5 deduced from optical spectra \citep{henry04}. It compares well with the spectrum of LP944-20, an M9V star that has $T_\mathrm{eff}$ between 2400\,K and 2600\,K \citep{basri00}. The spectrum was further fitted to synthetic spectra calculated with the general purpose stellar atmosphere code PHOENIX \citep{brott05}. The best fits were achieved with metallicities $-0.25<\log{Z}<0$ and 2600\,K $<$ $T_\mathrm{eff} < 2700$\,K. Hence, our observations are consistent with SCR 1845A being an M8.5 star, and we adopt an effective temperature of about 2600\,K.

Table\,\ref{tab:eqw} displays the atomic line equivalent widths of SCR 1845A calculated from its NIR spectrum. The line shape was interpolated using a spline fitting routine, and the continuum was derived by a linear fit to the data points outside the line. Uncertainties on the EWs were estimated from the minimum and maximum acceptable continuum values next to the lines. The measured equivalent widths are comparable to the ones of other stars of similar spectral type presented by \citet{cushing05}. They do not indicate extreme metallicity or gravity effects.

\begin{table}
\begin{minipage}[t]{\columnwidth}
\caption{Atomic line equivalent widths (\AA) of SCR 1845A and LP 944-20 for comparison. The KI equivalent widths are the sum of the doublets.}
\label{tab:eqw}
\centering
\renewcommand{\footnoterule}{}  
\begin{tabular}{llll}
\hline \hline
Object & KI (1.17$\mu$m)  &  KI (1.25$\mu$m) & NaI (2.21$\mu$m)  \\
\hline
SCR 1845A  & 15$\pm$3  &  24$\pm$3 & 2$\pm$1.2 \\
LP 944-20\footnote{\citet{cushing05}} & 14.4$\pm$0.6  &  9.4$\pm$0.6 &  2.7$\pm$0.3 \\
\hline
\end{tabular}
\end{minipage}
\end{table}

\subsection{Physical parameters of SCR 1845B}

We have explored the constraints on $T_\mathrm{eff}$, gravity ($\log_{10}{g}$), and object radius ($R$) of SCR 1845B due to its overall spectrum shape (in particular the K, J, and H bands flux ratios), its absolute flux levels, 
and the strength of the KI doublet in its J band. The extracted physical properties are provisional, but reasonable.  To fit the observations, first we generated a grid of atmospheric and spectral models for $T_\mathrm{eff}$ from 800 to 
1200\,K and for $\log_{10}{g}$ from 4.75 to 5.5 (cgs).  Next, those combinations of $T_\mathrm{eff}$ and $\log_{10}{g}$ that reproduced the shape of the observed spectrum, without consideration for the absolute flux level, were determined. Both low-temperature models with low gravities and high-temperatures models with high gravities can fit the shape of the spectrum. In this way, we derived the line in $T_\mathrm{eff}$--$\log_{10}{g}$ space for which our models fit the spectrum shape.  Then, assuming the evolutionary models of \citet{burrows97} and \citet{burrows06}, the {\it absolute} fluxes, which are functions of mass and age or, alternatively, of $T_\mathrm{eff}$ and $\log_{10}{g}$, were fit to the observed absolute flux of SCR 1845B, assuming a distance of 3.85\,pc.  The result was another line in $T_\mathrm{eff}$-$\log_{10}{g}$ space, this time one that fits the absolute flux constraint. Hence, using the spectrum shape and the flux separately, two independent lines in $T_\mathrm{eff}$--$\log_{10}{g}$ space were derived.  Naturally, the intersection of these two lines yields a single point that represents the $T_\mathrm{eff}$/$\log_{10}{g}$ pair that fits best overall. This point is at about $T_\mathrm{eff}=950$\,K and $\log_{10}{g}=5.1$ (cgs), which are the most probable values of those physical parameters within wide error bars. The procedure described above is similar to the one used by \citet{saumon06} to derive the physical properties of the T7.5 dwarf Gliese 570D.

Figure\,\ref{fig:jhkspectraB} displays the associated model spectrum together with the observed spectrum of SCR 1845B. This fit is for solar metallicity. Lower metallicity would decrease the relative K-band flux, which is also a decreasing function of gravity and an increasing function of temperature. Hence, lower metallicity would tend to shift the best fitting models to slightly higher values of $T_\mathrm{eff}$, or slightly lower values of gravity, all else being equal. 

Finally, we calculated profiles of the KI doublet at 1.25\,$\mu$m in the J band. These lines are barely discernible in the observed spectrum, but are present in most of our synthetic spectra. The model equivalent widths decrease towards 
lower $T_\mathrm{eff}$ and higher gravities. This is because the 1.25\,$\mu$m KI doublet requires high excitation temperatures.  Higher gravities place this high-temperature region at higher pressures and depths, thereby progressively burying it. So, if we were to weight our conclusions by the constraint imposed by the shallow KI doublet, we would shift our derived $T_\mathrm{eff}$ and gravities accordingly.  An interpretation of the weakness of these lines in the observed spectrum might also be that the potassium abundance is subsolar. However, given the otherwise good overall fit the metallicity itself is not appreciably subsolar.

Simultaneous correlated constraints on $T_\mathrm{eff}$ and radius of SCR 1845B can be obtained using the absolute flux level. Table\,\ref{tab:modelresults} lists the possible combinations, along with model radii. A consistent model 
for $T_\mathrm{eff}$ and radius is obtained at $T_\mathrm{eff} = 950-1000$\,K, and $R\approx0.63\times10^{10}$\,cm (0.88\,$R_\mathrm{J}$). This is also consistent with $\log_{10}{g} = 5.1-5.25$ and spectral type T6p.  Using the  \citet{burrows97} evolutionary models, we find mass/age pairs that are more or less consistent with the spectral fits of (41 $M_\mathrm{J}$, 1.8 Gyr), (51 $M_\mathrm{J}$, 3.1 Gyr), and (51.8 $M_\mathrm{J}$, 2.6 Gyr), with a very slight preference for the first pair.

\begin{table}
\caption{Parameters of SCR 1845B. First and second columns list possible combinations of the effective temperature ($T_\mathrm{eff}$) and radius ($R_{\mathrm{fit}}$) that fit the absolute flux from SCR 1845B.  Columns 3-5 list radii from evolutionary models ($R_{\mathrm{model}}$) that correspond to the particular combination of $T_\mathrm{eff}$ and $\log_{10}{g}$. Radii that best satisfy both constraints set by the absolute fluxes and evolutionary models are those with $R_{\mathrm{fit}} \approx R_{\mathrm{model}}$ and are printed in bold.}        
\label{tab:modelresults}      
\centering                          
\begin{tabular}{ccccc}        
\hline\hline                 
$T_\mathrm{eff}$ & $R_{\mathrm{fit}}$ (10$^{10}$\,cm) & & $R_{\mathrm{model}}$ (10$^{10}$\,cm) & \\
                 &                         & $\log{g}=5$             & $\log{g}=5.25$ & $\log{g}=5.1$ \\
\hline                        
 800 & 0.94 & 0.654 & 0.589 & \\      
 900 & 0.72 & 0.662 & 0.597 & \\
 950 & 0.63 & 0.666 & 0.601 & \textbf{0.639}\\
 1000& 0.58 & 0.670 & \textbf{0.604} & \\
 1100& 0.47 & 0.677 & 0.610 & \\
 1200& 0.40 & 0.684 & 0.616 & \\
\hline                                   
\end{tabular}
\end{table}

An approximate estimate of $T_\mathrm{eff}$ and $\log{g}$ can also be derived from spectral indices \citep{burgasser06b}. Applying this method using the indices given in Table\,\ref{tab:indices}, we find that the phase spaces of H$_2$O-J and K/H do not intersect assuming solar metallicity and staying inside the $\log{g}<5.5$ limit \citep{burrows97}. However, since both H$_2$O-J and K/H are more sensitive to metallicity than to surface gravity \citep{liu07}, only slightly subsolar metallicities are needed to find a valid combination of $T_\mathrm{eff}$ and $\log{g}$ which would reasonably well agree with the results from the direct fit of the model to the spectrum.

It is also interesting to compare SCR~1845B to the T6 dwarf $\epsilon$~Indi~Bb,which as part of the Ba/Bb binary system, is one of the two most nearby (3.626\,pc) brown dwarfs known today. Both NACO NIR photometry and H-band spectrum of $\epsilon$~Indi~Bb \citep{mccaughrean04} are very similar to our SCR~1845B data. The absolute J, H and Ks magnitudes of the two objects differ by less than 0.1 magnitudes from each other, with SCR~1845B being slightly brighter and bluer. The H-band spectra are virtually identical. Based on their data and assuming an age of 2\,Gyr, \citet{mccaughrean04} used the bolometric correction BC$_K$ presented by \citet[e.g.][]{golimowski04} and the \citet{burrows97} evolutionary models to derive a mass of 38 $M_\mathrm{J}$ for $\epsilon$~Indi~Bb. This result is a rather good match to our estimate of SCR 1845B's mass.

\section{Conclusions}

Following up the discovery of SCR~1845B, we have obtained J, H and Ks imaging and spectroscopy of the SCR 1845 binary. Given the large proper motion of SCR 1845, the relatively small (0.1\arcsec{}) change in relative position of the two stars between 2005 and 2006 confirms that the binary system is gravitationally bound. With only two data points available no further constraints on orbital parameters can be set for now.

Comparison with spectral standards and PHOENIX model fitting indicate spectral type M8.5 and $T_\mathrm{eff}\approx2600$\,K for the primary SCR 1845A as already derived before from the optical spectrum. A similar procedure using spectral standards and the University of Arizona evolutionary models for Brown dwarfs yields spectral type T6, $T_\mathrm{eff} \approx 950$\,K and $\log{g} \approx 5.1$ (cgs) for SCR 1845B. Considering also its measured luminosity, mass and age of SCR 1845B can be constrained to 40 to 50 Jupiter masses and 1.8 to 3.1 Gyr. SCR 1845B further exhibits strong K-band suppression and weak KI lines at 1.25\,$\mu$m both indicative of either high surface gravity or low metallicity. Hence, further dedicated observation of SCR 1845B might help us to obtain a better understanding of the complex interaction of surface gravity and metallicity in the formation of brown dwarf spectra.

The similarity between J, H, K absolute magnitudes and H-band spectrum of SCR 1845B and $\epsilon$~Indi~Bb is striking. Both T6 dwarfs have a main sequence companion and will have accurately measured masses (from measuring the orbital motion of SCR 1845AB and Eps Indi Ba-Bb) on time scales of a few years. Further high-resolution NIR spectroscopy and astrometry of SCR 1845B will therefore allow us to directly measure its mass and to constrain its physical parameters such as metallicity and age. This makes SCR 1845B one of the few precious objects known today that has the potential to provide the required benchmarks for testing evolutionary models.

\begin{acknowledgements}
      This research has made use of data products from the SuperCOSMOS Sky Surveys at the Wide Field Astronomy Unit of the Institute for Astronomy, University of Edinburgh, and from the Two Mircon All Sky Survey, a joint project of the University of Massachusetts and IPAC/Caltech, funded by NASA and the NSF. We thank Sandy Leggett for providing electronic M and T dwarf spectra through her Web site. We also thank Andreas Schweitzer for providing spectral fits to SCR 1845A using the PHOENIX code. WB acknowledges support by a Julian Schwinger fellowship through UCLA. AB and JB would like to acknowledge partial support from NASA grants NNG04GL22G and NNG05GG05G and through the NASA Astrobiology Institute under Cooperative Agreement No. CAN-02-OSS-02 issued through the Office of Space Science.
\end{acknowledgements}

\bibliographystyle{aa}

\bibliography{references}

%
%
%
%
%
%
%
%
%

\end{document}